\documentclass[onecolumn,english]{svjour3}
\usepackage[T1]{fontenc}
\usepackage[latin9]{inputenc}
\usepackage{array}
\usepackage{float}
\usepackage{amsmath}
\usepackage{amssymb}
\usepackage{graphicx}

\makeatletter

\providecommand{\tabularnewline}{\\}
\floatstyle{ruled}
\newfloat{algorithm}{tbp}{loa}
\providecommand{\algorithmname}{Algorithm}
\floatname{algorithm}{\protect\algorithmname}

\smartqed  % flush right qed marks, e.g. at end of proof

\makeatother

\makeatother

\usepackage{babel}

\makeatother

\usepackage{babel}
\begin{document}
\journalname{World Wide Web Journal}
\title{A Time Decoupling Approach for Studying Forum Dynamics}
\author{Andrey Kan	\and
	Jeffrey Chan	\and
	Conor Hayes	\and
	Bernie Hogan	\and
	James Bailey	\and
	Christopher Leckie
}
\institute{A. Kan (corresponding), J. Bailey, C. Leckie \at
	NICTA Victoria Research Laboratory, Department of Computer Science and Software Engineering, The University of Melbourne, Australia \\
	Tel.: +614 20 768 752\\
	\email{akan@csse.unimelb.edu.au} \\
	\and
	J. Chan, C. Hayes \at
	Digital Enterprise Research Institute, National University of Ireland, Galway, Ireland \\
	\and
	J. Chan\at
	Department of Computer Science and Software Engineering, The University of Melbourne, Australia \\
	\and
	B. Hogan \at
	Oxford Internet Institute, University of Oxford, United Kingdom
}
\maketitle
\begin{abstract}
Online forums are rich sources of information about user communication activity over time. Finding temporal patterns in online forum communication threads can advance our understanding of the dynamics of conversations. The main challenge of temporal analysis in this context is the complexity of forum data. There can be thousands of interacting users, who can be numerically described in many different ways. Moreover, user characteristics can evolve over time. We propose an approach that decouples temporal information about users into sequences of user events and inter-event times. We develop a new feature space to represent the event sequences as paths, and we model the distribution of the inter-event times. We study over $30,000$ users across four Internet forums, and discover novel patterns in user communication. We find that users tend to exhibit consistency over time. Furthermore, in our feature space, we observe regions that represent unlikely user behaviors. Finally, we show how to derive a numerical representation for each forum, and we then use this representation to derive a novel clustering of multiple forums.
\end{abstract}

\section{Introduction\label{S.1:intro}}

Online forums are popular websites that allow users to communicate
on diverse topics. These forums provide an extensive archive of conversation
records over time. Finding temporal patterns in these records can
advance our understanding of the dynamics of conversations. Furthermore,
forum analysis can facilitate solving practical tasks, such as predicting
the future behavior of users; identifying users whose behavior deviates
from that of the majority in a given forum; and providing an automated
forum categorization that can inform the structural design of forum
websites. Accordingly, the temporal analysis of online forum data
is of interest to different groups, including sociologists, businesses
that run customer support forums, and web designers.

The above-mentioned theoretical and practical advances can be advanced
by addressing the following research questions:
\begin{enumerate}
\item Are there any trends (e.g., consistency) in the temporal evolution
of users in online forums?
\item Is it possible to characterize the normative temporal behavior for
a given forum?
\item Can we quantify and categorize forums in terms of the temporal behavior
of users?
\end{enumerate}
The main challenges for the analysis of temporal forum data arise
from the complexity of the data. Communication records collected from
a medium-sized forum website (e.g., \emph{www.boards.ie} or \emph{forumbgz.ru})
for one year may comprise information about thousands of users, with
tens of thousands messages sent between different users. Furthermore,
there are dozens of features that can be used to describe users and
forums (e.g., number of initiated threads, number of connection ties,
number of threads in a forum, percentage of {}``answer people''
users \cite{chan2010decomposing,morzy2010analysis,viegas2004newsgroup,welser2007visualizing}).
It is not obvious which features to use, and how to define features
that evolve over time (e.g., should one consider the number of users'
posts per week or per month?)

Moreover, in order to have a computationally tractable online forum
analysis, one may need to select discrete time points for processing.
However, it is not obvious which time points to select. Finally, there
is the challenge of natural language processing. Different forums
can use different languages, and, for a given language, people may
mistype words, use slang and non-dictionary words.

Researchers have tried different approaches to forum data analysis,
such as focusing on user roles, discussion diversity and growth dynamics.
For example, both Viegas and Smith \cite{viegas2004newsgroup} and
Welser et al. \cite{welser2007visualizing} use visualization to derive
structural features of the roles of individual users, thus indicating
clear differences between consistent {}``answer people'' and those
preferring discussions. Lui and Baldwin \cite{lui2010classifying},
and Chan et al. \cite{chan2010decomposing} extend this work by categorizing
roles and exploring the composition of forums in terms of the diversity
of roles. Kumar et al. \cite{kumar2010dynamics} and Gonzalez-Bailon
et al. \cite{gonzalez2010structure} model growing trees of discussion
threads.

These studies tend to assume that users exhibit consistent behavior
over time, and that the users can be described using features that
are aggregated over the whole time span of the available data (e.g.,
total number of threads initiated by a user). Studies that look at
temporal patterns over time, instead of assuming consistency, tend
to be descriptive and difficult to scale. For example, \emph{AuthorLines}
\cite{viegas2004newsgroup} show the activity of individual users
over time, but with \emph{AuthorLines} it is hard to visualize the
evolution of the whole forum. \emph{To the best of our knowledge there
has been no forum representation that can be explicitly interpreted
in terms of the temporal behavior of individual users.}

To address these limitations, we propose a novel method for studying
forum dynamics. By forum dynamics we mean a set of changes in the
characteristics of the users participating in the forum over time.
We propose to decouple temporal information about users into sequences
of user events and inter-event times. Furthermore, we interpret a
sequence of user events as a path in a feature space; we represent
a forum with a set of paths; and we analyze forums from a geometrical
perspective.

In a given forum, we study user paths both individually and as a single
interdependent system. This makes our approach unique, because past
work on time sequences analysis mostly focuses on pairwise correlation
between parts of sequences (e.g., \cite{Mueen2009}). Moreover, our
analysis reveals interesting findings that were not previously reported.
This includes a roughly constant ability of users to attract responses
(consistency), and existence of user states that are very unlikely
to occur ({}``dead zones'' in a feature space). Finally, we propose
a forum visualization that allows us to obtain a succinct summary
of the behavior of users in forums.

In summary, the main contributions of our paper are as follows.
\begin{itemize}
\item A time decoupling approach for finding temporal patterns in online
forum data (Section \ref{S.4.1:tda}).
\item A new feature space to represent users' evolution (Section \ref{S.4.2:features}).
A proposed visualization of our feature space allows us to observe
users' characteristics that were not apparent from previous studies
(Section \ref{S.4.3:visual}).
\item A novel feature representation of forums (Sections \ref{S.6.1:deriving}).
This has enabled the discovery of a new clustering for one of the
previously studied datasets (Section \ref{S.6.2:categorization}).
\item Discovery of consistent engagement and dead zone patterns (Sections
\ref{S.5.1:consistent} and \ref{S.5.2:zones}) and the analysis of
inter-event times (Section \ref{S.5.3:times}).
\end{itemize}
Prior to giving a detailed description of our methods and results,
in the next two sections we first introduce related work and then
describe the datasets that we study.

\section{Related Work\label{S.2:related}}

In this section, we start with a review of approaches to study online
forum data (Table \ref{T:prev-work}). We then discuss work that studies
dynamic processes in related online media, such as online social networks
and weblogs.

\begin{table}[h]
\caption{\label{T:prev-work}Summary of related work on forums. There have
been different approaches to study forum data. Some studies focus
on individual users, while other studies look at the discussion threads
or forums.}

\centering{}%
\begin{tabular}{|>{\centering}p{0.21\textwidth}|>{\centering}p{0.13\textwidth}|>{\centering}p{0.32\textwidth}|>{\centering}p{0.2\textwidth}|}
\hline 
\textbf{References} & \textbf{Objects of study} & \textbf{Involved features} & \textbf{Representing user engagement over time}\tabularnewline
\hline 
Lui and Baldwin \cite{lui2010classifying}, Welser et al. \cite{welser2007visualizing} & users & structural features of reply graph such as proportion of low degree
neighbors, content-based features, etc. & not focused on changes over time\tabularnewline
\hline 
Chan et al. \cite{chan2010decomposing} & users, forums & number of posts per thread, ratio of in degree to out degree in reply
graph, etc. & not focused on changes over time\tabularnewline
\hline 
Viegas and Smith \cite{viegas2004newsgroup} & users, forums & number of initiated threads, number of threads where user participated,
etc. & real-time changes in statistics for a user\tabularnewline
\hline 
Xiong and Donath \cite{xiong1999peoplegarden} & users, forums & number of messages, time passed since posting, etc. & as a sequence of events (but not quantitative)\tabularnewline
\hline 
Kumar et al. \cite{kumar2010dynamics}, Wang et al. \cite{wang2010thread} & threads & degree distributions of discussion threads, content based features,
etc. & modeling the growths of conversation trees\tabularnewline
\hline 
Morzy \cite{morzy2010analysis} & micro-communities, forums & number of communities that sustain longer than certain amount of time,
number of communities per user, etc. & user engagement is not represented explicitly\tabularnewline
\hline 
this work & users, forums & number of posts and number of replies over time & as a sequence of events\tabularnewline
\hline 
\end{tabular}
\end{table}

Chan et al. \cite{chan2010decomposing} employ reply graphs, as well
as the structure of threads, to derive features for classifying users
into different roles (e.g., {}``taciturn'', {}``elitist''). In
the same work, they study the composition of forums in terms of user
roles. Lui and Baldwin \cite{lui2010classifying} classify users by
utilizing content based features in addition to structural features.
In both works, an implicit assumption is that user roles do not change
during the period of observation.

Welser et al. \cite{welser2007visualizing} derive features for individual
users by means of different visualization techniques. The features
are then used for automated identification of {}``answer people'',
a specific type of user. The automated identification is validated
by manually inspecting the contents of a selection of user posts.
In their study, Welser et al. state that {}``in many well-defined
circumstances, it is generally safe to assume that past behavior will
correlate with future behavior''. In our work, we present a framework
for assessing this assumption quantitatively by studying the temporal
evolution of users.

Viegas and Smith \cite{viegas2004newsgroup} also analyze temporal
evolution of users. They propose \emph{AuthorLines} for plotting the
number of threads started by a user, as well as the number of threads
where the user has participated for a particular week. It is then
easy to see, for example, how the number of these threads evolve over
subsequent weeks. However, it can be hard to compare the engagement
of many (e.g., $50$) users at once. Furthermore, Viegas and Smith
do not quantify patterns of user evolution. There have been other
proposed forum visualizations \cite{viegas2004newsgroup,xiong1999peoplegarden},
however these visualizations do not define numerical features for
describing and comparing the forums automatically.

Some authors study forums at the level of discussion threads or micro-communities
\cite{kumar2010dynamics,morzy2010analysis,wang2010thread}. Kumar
et al. \cite{kumar2010dynamics} present and evaluate a model of the
dynamics of conversations, where adjusting model parameters can lead
to the production of {}``bushy'' or {}``skinny'' thread trees.
While this model can be adapted to infer the evolution of individual
users, this adaptation is not straightforward and is not covered in
their paper. Furthermore, this model describes conversations on the
level of individual threads, whereas forums often consist of a number
of threads. Threads are modeled as trees whereas a forum is generally
a forest.

Furthermore, there has been a large body of literature that studies
processes in related media spaces such as Facebook, Twitter and weblogs.
Many researchers look at the dynamic processes in online media from
an information diffusion perspective. With this approach one models
the propagation of information among users in a similar manner to
modeling the propagation of infection in epidemiology \cite{Adar2005,Liao2011}.
Other approaches to studying evolving online social media include
the combined study of the diffusion and evolution of discussion topics
\cite{Lin2011}, modeling user interactions with dynamic graphs \cite{Viswanath2009},
and looking at the properties of time series of social interactions
\cite{Benevenuto}. Our work complements these earlier studies by
approaching online forums from the perspective of individual users
and the reciprocity of communication.

More generally, time series mining is an established field of data
mining \cite{Warrenliao2005}. Keogh et al. spell out an intuitive
idea that {}``representation of the data is the key to efficient
and effective solutions'' \cite{Keogh2003}. One of the novel contributions
of our work is the representation of the history of individual users
as two decoupled series: sequence of events and event timings.

In summary, forum data analysis is a field with growing research interest.
Clearly participating users constitute the essence of the forums,
and users' behavior eventually determines all the numerical features
of the forum data. However, there are several remaining open issues,
such as the lack of scalable methods for the representation and analysis
of the temporal evolution of individual users, and the lack of a numerical
representation for forums in terms of the evolution of users.

\section{Datasets\label{S.3:datasets}}

In our study, we use data from four Internet forums: \emph{Boards.ie},
\emph{SAP Community Network (SCN)}, \emph{TiddlyWiki}, and \emph{Ancestry.com}.
In total, we analyze records for over $30,000$ users. We observe
similar patterns in all datasets, and in the main text, we focus mainly
on results for \emph{Boards.ie}. We present results for the \emph{SCN},
\emph{TiddlyWiki}, and \emph{Ancestry.com} forums in Supplementary
Information 1.

\emph{Boards.ie} is a national Irish bulletin board that comprises
a variety of forums on different topics, from concert announcements,
to dedicated discussions about martial arts and politics. The records
for \emph{Boards.ie} were collected over a two year period (2006 and
2007).

We select nine \emph{Boards.ie} forums that represent a variety of
topics (Table \ref{T:forums}). In each forum, we select users that
had registered in $2006$. This selection allows us to observe the
evolution of users for at least one year. From the users registered
in 2006, in each forum, we randomly select $30\%$ of users.

\begin{table}[h]
\caption{\label{T:forums}Selected forums from \emph{Boards.ie} covering a
variety of topics (\#posts is the total number of posts made by all
users from the sample; \#replies is the total number of replies received
by all users from the sample)}

\centering{}%
\begin{tabular}{>{\centering}p{0.16\textwidth}>{\centering}p{0.13\textwidth}>{\centering}p{0.13\textwidth}>{\centering}p{0.13\textwidth}>{\centering}p{0.29\textwidth}}
\hline 
\textbf{Forum name} & \textbf{\#Users in sample}  & \textbf{\#Posts in sample}  & \textbf{\#Replies in sample}  & \textbf{Comments}\tabularnewline
\hline 
politics  & 264  & 3,867  & 3,761  & topical discussion\tabularnewline
soccer  & 73  & 4,352  & 4,215  & topical discussion\tabularnewline
poker  & 183  & 10,542  & 9,660  & topical discussion\tabularnewline
martial arts  & 156  & 5,444  & 4,911  & topical discussion\tabularnewline
personal  & 586  & 6,709  & 6,178  & discussion of personal issues\tabularnewline
accommodation  & 236  & 1,764  & 1,523  & classifieds and discussion\tabularnewline
gigs  & 345  & 2,141  & 1,715  & musical performance announcements and discussion\tabularnewline
weather  & 56  & 300  & 259  & forecasts and discussion\tabularnewline
development  & 107  & 679  & 590  & tech. questions and discussion\tabularnewline
\hline 
\end{tabular}
\end{table}

Sampling was used in order to speed up the computations. Note that
our sample is sufficiently large (there are around $2,000$ users
in the \emph{Boards.ie} sample) to draw conclusions. Furthermore,
we have studied the effect of sampling on two forums from \emph{Ancestry.com}
(as these forums have the largest numbers of users), and found that
taking a $30\%$ sample of users results in observations similar to
those made on the complete set of users. Details on our study of the
sampling effect are presented in Supplementary Information 2.

Note that the method presented in this paper operates at the level
of forums. The method will be mainly affected by the sizes of individual
forums rather than the size of a complete dataset. The size of the
forum can be measured, for example, using the average number of posts
per day. With respect to this measure, forums in our study appear
to be on the same scale with many other forums reported in the literature
\cite{kumar2010dynamics,morzy2010analysis}. This indicates that our
study is applicable to a large number of typical Internet forums.

We note that there can exist exceptionally large forums (e.g., with
more than $1000$ posts per day). On the other hand, we speculate
that there is a certain upper limit for the size of a forum. For example,
a forum with more than $1000$ posts per day may effectively become
unreadable for users, as it would be updated much faster than the
users can read. Such a forum might have a tendency to split into smaller
forums or to lose its participants, which would  bring the forum back
to a smaller size and a more balanced state.

Finally, it is worth mentioning that \emph{Boards.ie} was previously
studied by Chan et al. \cite{chan2010decomposing}. This enables us
to compare our results. We introduce our methods and results in the
next sections.

\section{Research Methods\label{S.4:methods}}

\begin{figure}[h]
\noindent \begin{centering}
\includegraphics{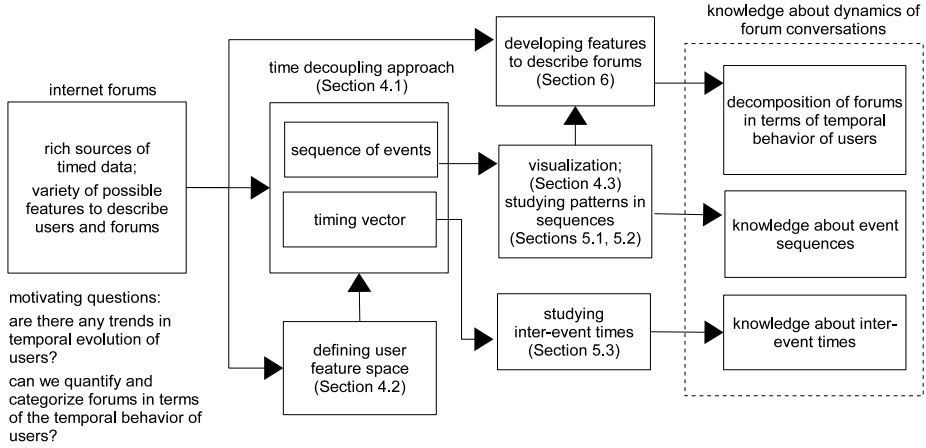}
\par\end{centering}

\caption{\label{F:plan}Outline of our research methodology. The goal is to
obtain novel knowledge about the temporal evolution of users (on the
right). In order to achieve this goal we use a time decoupling approach
and a feature construction to represent users with (i) paths in a
feature space, and (ii) timing vectors. We then study the paths and
vectors separately. Visualization of paths allows us to construct
features that can describe forums.}
\end{figure}

We outline our research methodology in Figure \ref{F:plan}. Our goals
are (i) to discover temporal patterns in user behavior, and (ii) to
use the obtained information to develop a numerical description of
forums in terms of the evolution of users. In order to achieve the
first goal, we seek a convenient representation of temporal data that
allows us to observe and quantify temporal patterns.

Previous work has shown that visualization is a powerful tool in forum
data analysis \cite{turner2005picturing,viegas2004newsgroup,welser2007visualizing}.
Therefore, we develop our representation of users with a novel time
decoupling approach, and select features in a way that allows us to
visualize user evolution. We start with a description of our time
decoupling approach in the next section.

\subsection{Time Decoupling Approach\label{S.4.1:tda}}

Consider forum data that spans over time window $[T_{0},T_{1}]$,
for example from January $2006$ till December $2008.$ Let $U$ be
the set of all users in the forum, and let user $u\in U$ be registered
in the forum at time $T_{0}(u)\geq T_{0}$. A user feature is some
quantity that characterizes the user over time, for example the number
of posts the user $u$ has made by the time $t$.
\begin{definition}
\label{D:user-feat}A \textbf{user feature} $f(u,t)$ is a function
$U\times[T_{0}(u),T_{1}]\rightarrow\mathbb{R}$.
\end{definition}
Note that we assume that the feature values are real valued. In practice,
many widely used features (e.g., degree of a user in the reply graph,
number of posts per thread) can be mapped to real values.

At any time point $t\in[T_{0}(u),T_{1}]$, the user can be numerically
described with a set of features $F(u,t)=\{f_{1}(u,t),\dots,f_{n}(u,t)\}$.
We note that it is generally safe to assume that every user feature
$f_{i}\in F$ is a piecewise-constant function. This is the case with
many features used in practice, e.g., the number of user posts only
changes at certain time points, rather than continuously. Under this
assumption, we can describe a user with two finite sets defined below.
\begin{definition}
Given a set of piecewise-constant user features $f_{i}(u,t)$, $i=1,\ldots,n$,
a \textbf{timing vector} of user $u$ is the ordered set $T_{event}(u)=\{T_{0}(u),t_{1},\ldots,t_{L}:T_{0}(u)\leq t_{1}<\ldots<t_{L}\leq T_{1}\}$,
where $t_{i}$ is a time point at which at least one of the features
$f_{i}(u,t)$ changes its value. A \textbf{user path} of user $u$
is the set $\mathbb{P}(u)=\{F(u,t):t\in T_{event}(u)\}$. A \textbf{path
length} is the number $L(u)=|T_{event}(u)|-1$.
\end{definition}
Note that the user path can be viewed as a sequence of events listed
in chronological order, where an event is a change in feature values.
We suggest to separate the study of the user paths and the timing
vectors. We expect that prominent patterns in user evolution are preserved
in user paths with the order of events. On the other hand, time stamps
are more noisy, in the sense that they are likely to be influenced
by various real world factors that are either not interesting for
the study, or very hard to account for (e.g., sickness of a user,
weather conditions). Therefore, combining the time stamps with the
user paths might obfuscate important patterns.

While this framework is general, in a sense that it can incorporate
any number of features, it is particularly convenient to consider
a two-feature representation, as it allows us to easily plot the user
paths. We propose to represent users with two new features: $f_{1}(u,t)$
is the number of posts user $u$ has made since registration to time
$t$ inclusive, and $f_{2}(u,t)$ is the number of replies user $u$
has received since registration to time $t$ inclusive.

\begin{figure}[h]
\noindent \begin{centering}
\includegraphics{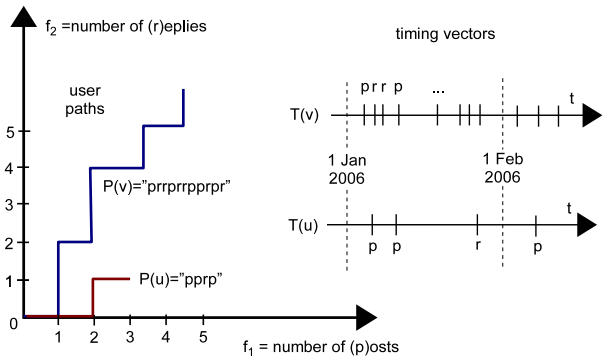} 
\par\end{centering}

\caption{\label{F:defs}Two sample user paths and the corresponding timing
vectors. Two features are used to represent the users. The order of
changes in feature values is captured by user paths. The timing of
the events is recorded in timing vectors.}
\end{figure}

Figure \ref{F:defs} shows two sample user paths and the corresponding
timing vectors. Consider path $\mathbb{P}(u)$ in the figure. At the
registration time, user $u$ has no posts and no replies, and therefore
the path starts from $(0,0)$. User $u$ then posts twice, receives
a reply, and posts again, thus $\mathbb{P}(u)$ goes to $(1,0)$,
$(2,0)$, $(2,1)$, and $(3,1)$. There is no more recorded activity
for user $u$, and so the path stops. We assume that at each point
in time, there can be only one post or one reply. The two possible
events in this case are incrementing the number of posts (p) and incrementing
the number of replies (r). Note that users $u$ and $v$ do not necessarily
reply to each other. We discuss our choice of user features in the
next section.

\subsection{User Feature Space\label{S.4.2:features}}

Recall that we aim to represent a user with an interpretable feature
set that can be visualized. There are a large variety of previously
proposed features to describe users \cite{chan2010decomposing,viegas2004newsgroup,welser2007visualizing},
and there are a number of methods for automated feature selection.
Broadly relevant feature selection methods can be divided in two groups:
feature transformation (such as well known principal component analysis),
and unsupervised feature selection (see for example \cite{boutsidis2009unsupervised,somol2010efficient}).

Unfortunately, both groups of methods were developed for a setup that
differs from the one that we observe in our dataset. Specifically,
previous work considers a finite set of objects, where each object
is described by a set of numbers. Here, for a given object, \emph{a
feature is a number}. In our case, we are interested in the temporal
evolution of users. We have a finite set of users, but each user is
described by a set of functions of time. We have that for a given
user, \emph{a feature is a function} (see Definition \ref{D:user-feat}).

Furthermore, previously proposed features are defined as numbers.
For example, the number of initiated threads \cite{chan2010decomposing}
is counted over the whole time span of the dataset and does not vary
with time. We therefore construct new time varying features, instead
of selecting from previous features.

As we stated in the previous section, we propose two new user features:
the number of posts a user has made since the registration to time
$t$ (inclusive), and the number of replies the user has received
since the registration to time $t$ (inclusive). The two features
comprise our feature space. In this space, a user path is constructed
as follows. The path starts from point $(0,0)$. Whenever a user makes
a post, the path goes one unit to the right. Whenever a user receives
a reply the path goes up one unit. We count a self-reply only once
as a post, and we assume that at one time point there can be only
one post or one reply. Note that the path always starts with a post,
as there cannot be replies when there are no posts.

We would like to clarify that a post always belongs to exactly one
thread. Within the thread, a post can be the starting post for the
thread or it can be a reply to any previous post in the same thread
(those posts, can in turn be replies to earlier posts). In some forum
settings, one post can even be a reply to more than one previous post
in the thread \cite{Kim2010}. In all these cases, the post moves
its author one unit to the right.

We now discuss the advantages and limitations of the proposed feature
space.

\subsubsection*{Advantages of Our Feature Space}

\begin{figure}[h]
\noindent \begin{centering}
\includegraphics{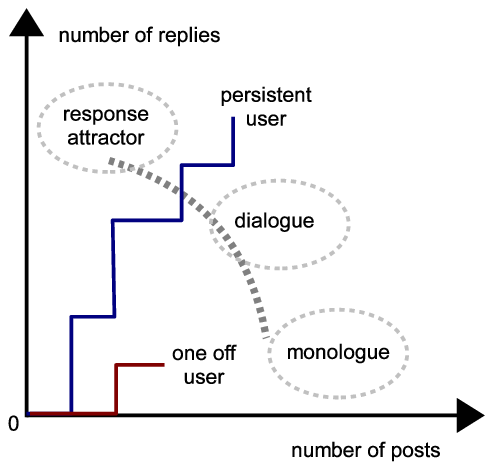} 
\par\end{centering}

\caption{\label{F:interpretation}Sample user paths are shown in a 2D space
that comprises two features: the number of user posts and the number
of replies received by the user. In this feature space, the path length
represents persistence and the slope represents reciprocity of communication.
For example, a path closer to the x-axis contains longer consecutive
posting sequences. Such a path can be categorized as a {}``monologue''.
In contrast, a {}``dialogue'' is a more balanced path where posts
tend to be followed by replies.}
\end{figure}

First, our feature space can be visualized and interpreted in terms
of persistence and reciprocity of communication (Figure \ref{F:interpretation}).
We believe that a user path represents a solid operationalization
of generalized reciprocity at the ego level with something coming
back to ego from many different sources \cite{plickert2007s}. Ego-level
generalized reciprocity has two qualities that may be expressed as
quantity and focus. Quantity, or how much feedback does one get for
a given post (on average), can be measured by the replies to posts
ratio of the user path. Focus is instead a measure of whether the
community is being reciprocal to the individual or is focused on particular
posts. If the individual path has a high level of deviation from the
straight line, then we can consider the community to be more focused
on the posts than the individual. If the deviation is low, that can
indicate that the community is oriented towards the individual, that
is, the community is interacting more with the individual than with
specific posts.

Second, these features are among the most straightforward measures
of a user's activity. Furthermore, these features are independent
of the language of the forum.

Third, we find that, in our datasets, the user paths in this feature
space have strong relations with some widely used structural and temporal
properties of users. A \emph{reply graph} \cite{welser2007visualizing}
is a popular structure, where vertices represent users, and directed
edges represent replies made during the time span of the dataset.
We find a strong correlation between the path lengths and the total
degree of users in the reply graph. In all of our forums we observe
Spearman's $\rho>0.9$ and $p<0.001$. We use Spearman's $\rho$ because
the distribution of path lengths is skewed.

Furthermore, let a \emph{user's up-time} be the time lapsed between
the first and last recorded posts made by a user. We find a strong
correlation between the path lengths and the up-times. In every forum
we observe Spearman's $\rho>0.7$ and $p<0.001$. The observed correlations
match the intuition, e.g., one can expect users with a large number
of posts to have a large number of correspondents. Bird et al. have
made a similar observation for an email dataset \cite{bird2006mining}.

Finally, we note that the chosen features -- numbers of posts and
replies -- have meaningful counterparts in other domains. Examples
include numbers of sent and received emails, outgoing and incoming
calls or text messages, or even, more generally, numbers of stimuli
and actions.

\subsubsection*{Limitations of Our Feature Space}

We do not take into account the variation among alters who respond
to ego. Indeed, ego might get a large quantity of reciprocity from
many different alters (or responders), or ego might have one single
fan that will consistently reply to ego\textquoteright{}s posts. Once
one begins to consider the variation among alters in this structure,
the outcome is more complex. This approach has already been taken
by Welser et al. \cite{welser2007visualizing}, as they looked at
whether individuals received replies from alters of low or high degree.
We do not dismiss this approach. Rather, we are looking for an efficient
metric that can summarize many of these more complex features in simpler
terms.

The defined features rely on the availability of reply relations between
messages. In some forums this information is recorded explicitly (e.g.,
this information was available in our datasets). In the case when
the reply relations are not available, one can use previously proposed
algorithms to reconstruct the reply relationships \cite{aumayr2011reconstruction,petrovcicposting}.

Visualization is a powerful tool, but it tends to limit the number
of dimensions that can be shown. In this paper, we do not consider
a large number of features. However, we have demonstrated that even
a small number of carefully constructed features can provide useful
insights into forum behavior. Moreover, recall that the user paths
can also be represented as sequences. Sequence pattern mining is a
well developed field (see for example \cite{loekito2010binary} and
the references therein). Therefore, in order to identify patterns,
one can use a sequence mining technique as an alternative to the visualization.
In this case, a larger number of features can be used. We leave this
investigation for future work.\\

Despite the potential limitations listed above, our feature space
allows us to discover novel patterns (e.g., consistency in user behavior)
that were not apparent from the previous user representations. Furthermore,
our feature space has inspired us to develop a new numerical description
of forums. We first summarize our visualization technique, and then
present our results in subsequent sections.

\subsection{Visualization\label{S.4.3:visual}}

\begin{figure}[h]
\begin{centering}
\includegraphics[scale=0.6]{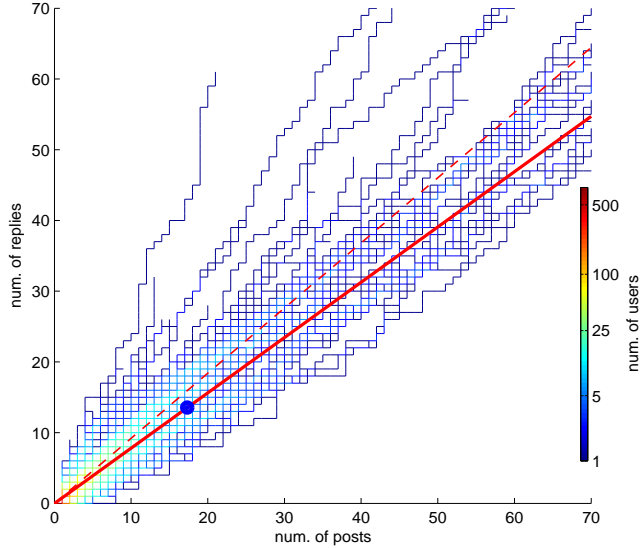} 
\par\end{centering}

\caption{\label{F:oneforum}{}``Personal'' forum from \emph{Boards.ie} is
visualized with a set of superimposed user paths. At each point, color
encodes the number of user paths going through that point. Together
the user paths show the trends in conversations specific to the forum.
The dashed red line is the baseline that shows the replies to posts
ratio. In this forum many paths significantly deviate from the baseline
and the mean slope of the forum (solid red line) goes below the baseline.}
\end{figure}

We propose to visualize forums with superimposed user paths (Figure
\ref{F:oneforum}). The paths can cross each other and overlap, and
at each point in the plot, color encodes the number of user paths
going through that point. The user paths are aligned with respect
to the beginning of the users' activity, not in real time. For each
path, segment $(0,0)$ -- $(1,0)$ corresponds to the first post made
by a user since registration, but this first post is made at different
times for different users. Such an alignment allows us to compare
users as they evolve in the forum.

The overall slope of a forum reflects the reciprocity of communication
in the forum. As a reference, we also draw the baseline degree line,
which represents the replies to posts ratio in the forum. The mean
length of paths represents the persistence of user engagement in a
forum, and the spread of paths in space represents the variety of
users. Superimposing paths allows us to reveal the trends in conversations
common to users in the given forum.

An advantage of our visualization technique is the scalability in
terms of the number of users that can be shown. In principle, this
number is not limited. The analogy here is plotting histograms or
scatter plots that are capable of showing prominent patterns of populations
of arbitrary sizes.

At the same time, we emphasize that visualization is not essential
for our numerical analysis methods that we present in the next few
sections.

\section{Patterns in Forum Data\label{S.5:patterns}}

Recall our first and second research questions stated in the introduction.
\begin{enumerate}
\item Are there any trends (e.g., consistency) in the temporal evolution
of users?
\item Is there a normative temporal behavior for a given forum?
\end{enumerate}
In this section, we address these questions by studying the user paths
and timing vectors.

\subsection{Pattern 1. Consistent Behavior\label{S.5.1:consistent}}

We plot all forums using our visualization method. The main trend
that we observe in the plots is that user paths can be approximated
with straight lines at different angles to the x-axis. We quantify
this linear dependency by computing Pearson's correlation coefficient
$r$ between the x and y coordinates of points for individual user
paths. That is, for each path we compute the correlation coefficient
$r$ separately. For the \emph{Boards.ie}, \emph{SCN}, \emph{TiddlyWiki},
and \emph{Ancestry.com}%
\footnote{In Section \ref{S.5.1:consistent}, for \emph{Ancestry.com} dataset
all statistics were obtained from the entire set of users, not a sample.%
} datasets the average $r$ across all users in the corresponding dataset
is $0.959$, $0.943$, $0.937$ and $0.921$ respectively. The computed
correlations are statistically significant ($p<0.01$). Here we only
compute correlation for users whose path contains sufficiently many
points. We consider users with path lengths of at least $10$. The
selection leaves us with $4070$ paths across all forums.

We conclude that users tend to exhibit consistency in communication
over time. We further investigate this consistency by modeling the
user paths.

\subsubsection*{Modeling User Paths}

We note that the user paths can be represented with strings comprising
posts (symbol 'p') and replies (symbol 'r'). Therefore, as a null
model for the user paths we use the well-known \emph{coin toss model}
(also known as a Bernoulli process). In this model, symbols 'p' and
'r' are generated sequentially. Our model always starts by producing
the symbol 'p', because the real paths always start with a post. Then
at each step, the symbol 'p' is chosen with a fixed probability $P_{post}$
that describes the ability of a user to elicit replies. Larger values
for $P_{post}$ result in a larger proportion of posts in the path,
and this corresponds to a weaker ability to elicit replies. For each
user we empirically estimate $P_{post}$ as $P_{post}\approx\#posts/(\#posts+\#replies)$.
We evaluate this null model by looking at the distribution of post
runs in the data.

\begin{figure}[h]
\begin{centering}
\includegraphics[scale=0.5]{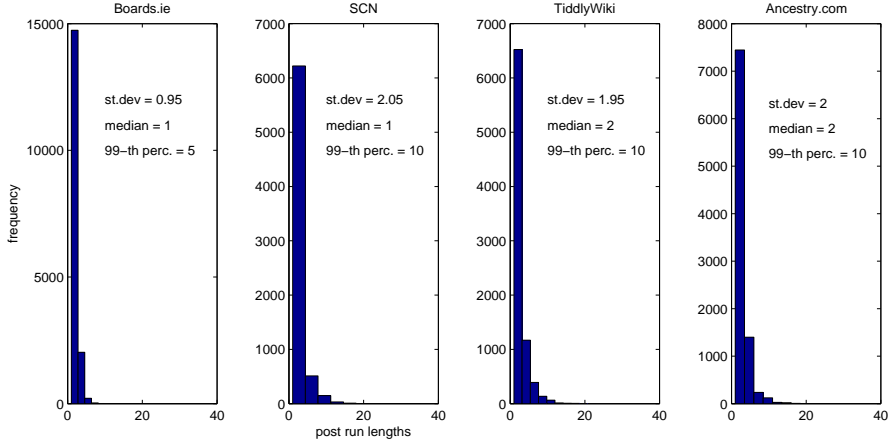}
\par\end{centering}

\caption{\label{F:post-runs}Empirical distributions of post runs aggregated
over all tested users in different datasets. Post runs and their standard
deviations tend to be short compared to the lengths of paths (over
$20$). This can be a consequence of consistency in user behavior.}
\end{figure}

By analogy with runs in binary strings, we define a post run as follows.
\begin{definition}
Consider a user path $\mathbb{P}$ represented with a string of posts
(symbol 'p') and replies (symbol 'r'). Let the corresponding extended
path be the string $\mathbb{P}^{\prime}=r\mathbb{P}r$. In the path
$\mathbb{P}$, a \textbf{post run} is a substring in a form $rp^{+}r$
taken from the corresponding extended path $\mathbb{P}^{\prime}$.
Here $p^{+}$ denotes an arbitrary positive number of posts. A \textbf{post
run length} is the number of posts in the post run. Where unambiguous,
we refer to the post run length as a post run.
\end{definition}
We now look at the distribution of post runs in individual user paths.
We select users with a path length of at least $20$, to ensure that
there are several post runs in each path. Figure \ref{F:post-runs}
shows the empirical distributions of post runs aggregated over all
tested users in different datasets. We find that post runs and their
standard deviations tend to be small compared to the lengths of paths
(over $20$). This can be a consequence of consistency in user behavior.
We are interested in whether the coin toss model is able to reproduce
the empirical distribution of post runs for individual users that
is present in the data.

For each tested user, we estimate $P_{post}$ and record the path
length. We then generate a path of this length using the coin toss
model with the estimated parameter $P_{post}$. For each user, we
then compare the empirical distribution of post runs and the generated
distribution using the Kolmogorov-Smirnov (KS) test at a $5\%$ significance
level. A user \emph{passes the test} if the KS test fails to reject
the null hypothesis that the distributions are different. The number
of users that pass the test is shown in Table \ref{T:models}. We
conclude that in general, the coin toss model adequately renders the
user paths.

\begin{table}[h]
\caption{\label{T:models}Results of models evaluation. {}``Users Tested''
shows the number of users with a path length of at least $20$ for
each of the datasets. Both models are capable of recovering the distributions
of post runs for the majority of users. However, in the \emph{Boards.ie}
dataset about $27\%$ of users fail the test.}

\centering{}%
\begin{tabular}{c>{\centering}p{0.1\textwidth}>{\centering}p{0.12\textwidth}>{\centering}p{0.12\textwidth}c}
\hline 
\textbf{Dataset}  & \textbf{Users Tested}  & \textbf{Users Passed (Coin)}  & \textbf{Users Passed (Sticking)}  & $P_{harsh}$\tabularnewline
\hline 
\emph{Boards.ie}  & 186 & 135 & 178 & 0.8\tabularnewline
\emph{SCN} & 112 & 101 & 109 & 0.8\tabularnewline
\emph{TiddlyWiki} & 115 & 97 & 93 & 0.6\tabularnewline
\emph{Ancestry.com} & 136 & 120 & 134 & 0.7\tabularnewline
\hline 
\end{tabular}
\end{table}

At the same time, we find that in the \emph{Boards.ie} dataset about
$27\%$ of users fail the test. We hypothesize that in this dataset,
users tend to be more consistent (i.e., follow a straight line more
closely, as explained below) than the random users generated by the
coin toss model.

We therefore propose another model for the generation of user paths.
We call this model the \emph{sticking model}, because in this model
users are stuck to a straight line by penalizing deviations. The slope
of the straight line is determined by the parameter $P_{post}$ that
is estimated for each user as $P_{post}\approx\#posts/(\#posts+\#replies)$.
The model is summarized in Algorithm \ref{A:stick}. In the model,
as soon as the generated path deviates from the straight line, a harsh
probability $P_{harsh}$ (e.g., $P_{harsh}=0.8$) forces the model
to generate the next symbol in a way such that the balance is restored.
Note that the sticking model can only be reduced to the coin toss
model in the case when $P_{harsh}=0.5$, which corresponds to the
coin toss model with $P_{post}=0.5$.

\begin{algorithm}[h]
\caption{\label{A:stick}Sticking model}

\noindent \textbf{Input:} path length $L$, user characteristic $P_{post}$,
dataset characteristic $P_{harsh}$;

\noindent \textbf{Output:} string $\mathbb{S}$;

\let\oldenumerate=\enumerate
\def\enumerate{\oldenumerate%
\setlength{\itemsep}{0pt}\setlength{\parsep}{0pt}}%
\renewcommand{\labelenumi}{\scriptsize{\arabic{enumi}:}}
\begin{enumerate}
\item $\mathbb{S}\Leftarrow'p'$; // real paths always start with posts
\item $currProp\Leftarrow1$; // proportion of posts in the string
\item \textbf{for} each symbol $currLength=2...L$
\item ~~~~$P_{curr}=\begin{cases}
P_{harsh}, & P_{post}>currProp\\
1-P_{harsh}, & P_{post}<currProp\\
0.5, & P_{post}=currProp
\end{cases}$;
\item ~~~~$currSymb=random(P_{curr},'p','r')$; // select $'p'$ with
probability $P_{curr}$;
\item ~~~~$\mathbb{S}\Leftarrow\mathbb{S}+currSymb$; // append the
string
\item ~~~~$currProp\Leftarrow\#(posts\ in\ \mathbb{S})/currLength$;
// update the proportion
\item \textbf{end}
\item \textbf{return} $\mathbb{S}$;\end{enumerate}
\end{algorithm}

We evaluate the sticking model on each user individually, similar
to our evaluation of the coin toss model. To simplify and speed-up
the calculations, we use the same $P_{harsh}$ for all users in the
same dataset. For each dataset, we find the value that maximizes the
number of users that pass the test, using grid search. The results
are presented in Table \ref{T:models}. We find that in general the
sticking model explains the empirical data better than the coin toss
model. The difference in models is especially prominent in the \emph{Boards.ie}
dataset. Note that this dataset also has the strongest mean correlation
coefficient (Section \ref{S.5.1:consistent}), which indicates that
users in this dataset have the strongest tendency to adhere to straight
lines.

From our modeling experience, we conclude that the consistency of
a user can generally be explained by an inherent user property. We
expressed this property with a constant parameter $P_{post}$. At
the same time, small deviations from the parameter are possible, and
the amount of this deviation varies in different forums. We can express
this freedom to vary by using the parameter $P_{harsh}$. For example,
in \emph{Boards.ie} users exhibit a stronger consistency than one
would expect for a random user modeled with the coin toss model.

\subsection{Pattern 2. Dead Zones\label{S.5.2:zones}}

Another prominent pattern that we observe in our visualizations is
that the user paths are not equally distributed in the feature space.
We formalize this pattern with the concept of a dead zone.
\begin{definition}
A \textbf{dead zone} is the set of points $(x,y)$ with integer coordinates,
such that the probability of observing a user in each point $(x,y)$
is less than some threshold value $P_{dz}$.
\end{definition}
In order to calculate the empirical probability of user locations,
we treat points from all paths in a forum as a single set. We then
use a kernel density estimation method by Botev et al. \cite{botev2010kernel}.
This method performs an automated selection of kernel bandwidth. However,
with the automatically selected bandwidth, the empirical density has
a grid patterning, as all our points have integer coordinates. Therefore,
we smooth the kernel by increasing the bandwidth in both dimensions
to $2$. We limit the region for the density estimation to the square
that has its lower left corner at point $(0,0)$, and its side equals
to the largest coordinate observed among forum points along either
the x or y axis. Finally, we set the threshold $P_{dz}$ to the $5$-th
percentile of the probabilities of all observed user locations in
the forum.

\begin{figure}[h]
\begin{centering}
\includegraphics[scale=0.6]{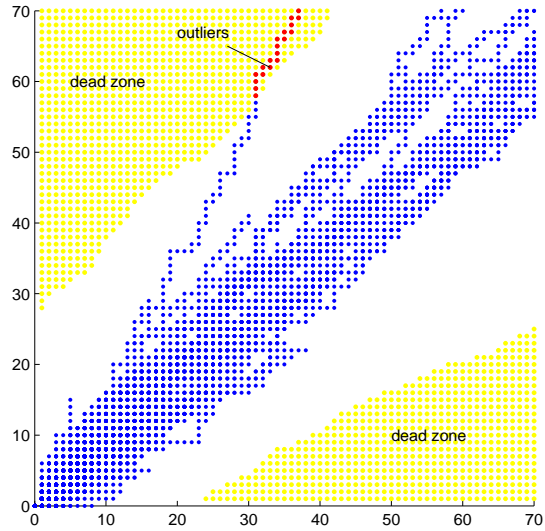}
\par\end{centering}

\caption{\label{F:zones}Dead zones (shown in yellow) for the politics forum
from \emph{Boards.ie}. The blue dots between the two dead zones show
empirically observed user locations. Red dots show the outlier locations
that can indicate an 'abnormal' user behavior. We observe similar
dead zones in all our forums.}
\end{figure}

The dead zones for the politics forum from \emph{Boards.ie} are shown
in Figure \ref{F:zones}. In general, the probability of observing
a user in the upper-left or lower-right corners of the plot is low.
We observe similar dead zones in all our forums.

The dead zones can be considered as a formal definition of a normative
user behavior for a given forum, and user posts found in the dead
zones can indicate an 'abnormal' user behavior (see Figure \ref{F:zones}).
For example, we would expect to find spammers in the zone along the
x-axis, and we would expect a user with highly provocative behavior
(attracting many replies) along the y-axis.

\subsection{Patterns in Timing Vectors\label{S.5.3:times}}

\begin{figure}[h]
\begin{centering}
\includegraphics[scale=0.6]{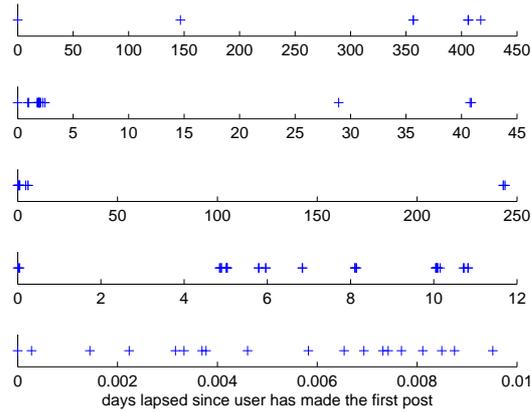} 
\par\end{centering}

\caption{\label{F:times}Timing of events of five randomly selected users.
The x-axis shows the number of days lapsed since the corresponding
user has made the first post. Note the different scales on the x-axes.}
\end{figure}

We now study the second part of our user data model, the timing vectors.
We start with plotting events of random users along the time line
(Figure \ref{F:times}). In the plot, the x-axis shows the number
of days lapsed since the user has made the first post. A cross mark
indicates an event (a post or a reply). Although users can have very
different time intervals (note the different scales on the x-axes),
we note that in general users exhibit bursty behavior: events tend
to group together in time.

\begin{figure}[h]
\begin{centering}
\includegraphics[scale=0.5]{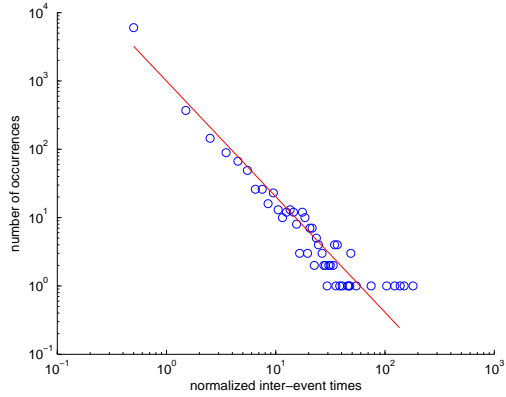} 
\par\end{centering}

\caption{\label{F:dt-distr}The distribution of normalized inter-event times
of users in politics forum. The plot shows about $7,000$ points in
a log-log scale.}
\end{figure}

We confirm this grouping pattern by looking at the distribution of
inter-event times. We hypothesize that such a distribution is heavy
tailed, with high probability values for a small range of short times
and low values for all other times, as short times correspond to times
between events in a burst, and long times correspond to times between
bursts. This supposition is confirmed when we look at empirical distributions
of the inter-event times collected for individual users in our forums.
For each user (with at least $10$ events) we collect a set of inter-event
times and divide the numbers by the mean value. This results in normalization
of the inter-event times across users. We then combine the sets of
the normalized inter-event times for all users in a forum and fit
a power-law curve. A representative result of such a fit is shown
in Figure \ref{F:dt-distr}. We observe a similar distribution in
every forum that we study with the fitted characteristic exponents
of about $-1.7$.

This bursty behavior is a well-studied phenomenon in domains such
as email communications \cite{barabasi2005origin}, and there have
been effective models defined for this behavior \cite{malmgren2008poissonian}.
The model of Malmgren et al. \cite{malmgren2008poissonian} for inter-event
times complements the models that we introduce in Section \ref{S.5.1:consistent}
for user paths. Together these models summarize a user's behavior
over time.

\begin{figure}[h]
\begin{centering}
\includegraphics[scale=0.6]{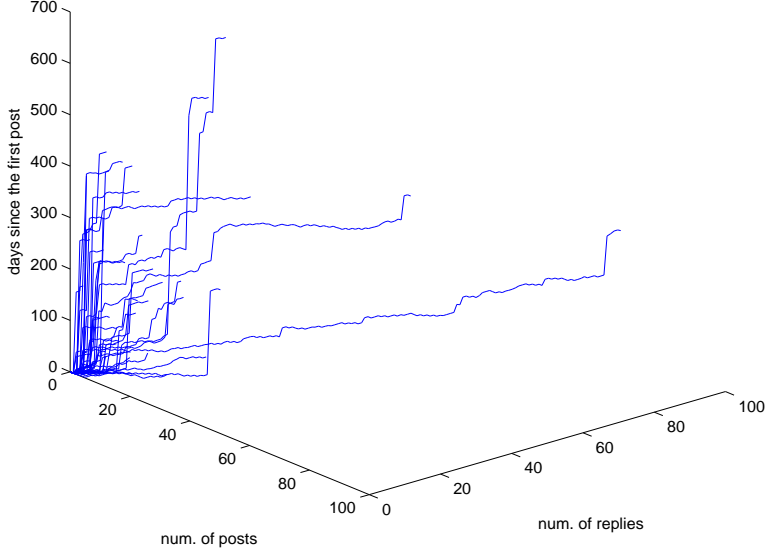} 
\par\end{centering}

\caption{\label{F:3d-plot}Plotting user paths in 3D with the numbers of posts
and replies and timings of events. The consistency in user behavior
is not apparent anymore. There is a staircase pattern in the temporal
dimension, but this is a consequence of the bursty behavior that we
discussed in this section.}
\end{figure}

There can be cases when studying user paths together with event timings
is important. However, we note that combining the two parts can obfuscate
otherwise prominent patterns (see Figure \ref{F:3d-plot}). This is
consistent with our motivation for using a time decoupling approach.

\section{Describing Forums\label{S.6:macro}}

\begin{figure}[h]
\begin{centering}
\includegraphics[scale=0.5]{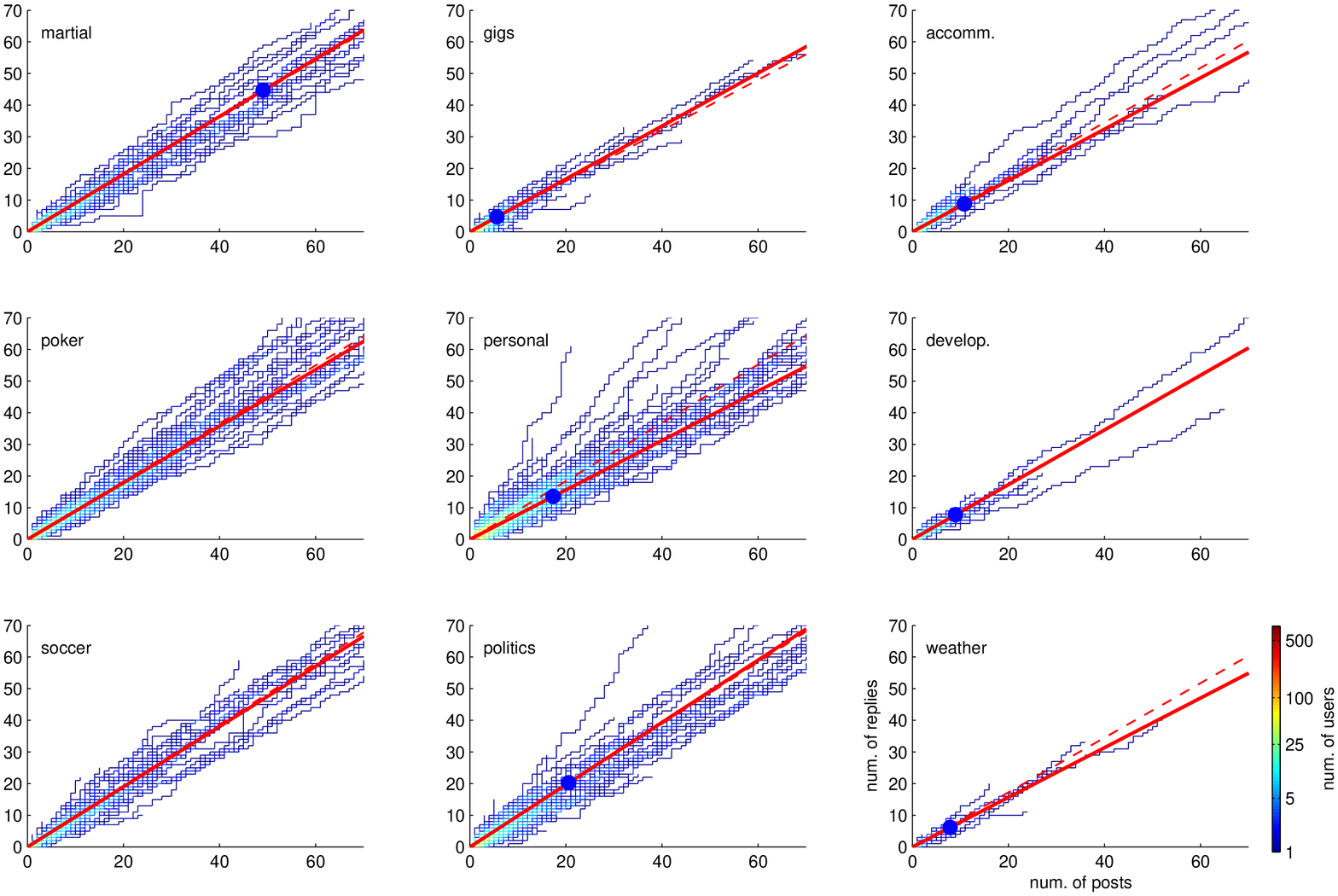} 
\par\end{centering}

\caption{\label{F:allforums}Visualization of forums from \emph{Boards.ie}.
Each line is a user path. At each point, the color encodes the number
of paths going through the point. The dashed red line is the baseline
that reflects replies to posts ratio, the solid red line shows the
average slope of the forum. The blue dot shows the average length
of user paths for the forum measured along the slope line (for the
soccer and poker forums the average length is above $50$). Differences
in forum appearance (different mean length or slope) indicate differences
in the nature of communication in the corresponding forums.}
\end{figure}

In this section, we address our third research question: {}``Can
we quantify and categorize forums in terms of the temporal behavior
of users?'' We start with the development of a concise numerical
description of forum visualizations such as that shown in Figure \ref{F:allforums}
(see also Supplementary Information 3).

\begin{definition}
Let a \textbf{forum archive} $\mathbb{A}$ denote a set of user paths
$\mathbb{A}=\{\mathbb{P}(u)\}$ observed in a forum within the time
span of the dataset. A \textbf{forum feature} is a function defined
on the forum archives $f:\mathbb{A}^{*}\rightarrow\mathbb{R}$, where
$\mathbb{A}^{*}$ is the set of all possible forum archives.
\end{definition}

\subsection{Deriving Forum Features\label{S.6.1:deriving}}

One possible way of deriving the forum features is to fit each path
independently with a straight line and then use the mean length and
slope of the paths. However, using the mean slope is not straightforward.
Paths with different lengths have different influence on the overall
shape of the forum. Further, different paths have different RMSE when
fit with a line. It is not obvious how to weight the paths appropriately.
Therefore, we take a different approach. We treat points from all
paths as a single set, and define forum features using this set.

\subsubsection*{Forum Slope and Spread}

Our first feature is called the slope feature $\tilde{\beta}$. This
feature is derived as follows. We fit a straight line to all points
in the forum, and compute the angle between the fitted line and the
x-axis. Recall that in a forum, for each user $u\in U$, the path
with length $L$ is the set of points $\mathbb{P}(u)=\{(0,0),(x_{u1},y_{u1}),\ldots,(x_{uL},y_{uL}),\}$.
Let $\mathbb{Q}$ be the union of points from all paths $\mathbb{Q}=\cup_{u\in U}\mathbb{P}(u)$.
The RMSE of fitting a line to the forum is defined as 
\begin{equation}
ERR(\beta)=\sqrt{\frac{\sum_{(x,y)\in\mathbb{Q}}(y-\beta\cdot x)^{2}}{|\mathbb{Q}|}}.\label{E:err}
\end{equation}
Here $\beta$ is the parameter of the line that is estimated by the
fitting process. Note that we force the fitted line to pass through
the point $(0,0)$. This allows us to reduce the number of fitting
parameters to one ($\beta$). Also note that points from different
paths are considered as different elements, e.g., $\mathbb{Q}$ may
contain several points $(1,1)$ from different users. This introduces
an implicit weighting of user paths, because paths that have more
points have more influence on the statistic. In every forum that we
study, the majority of users have short paths (one or two posts).
Without the weighting these short paths will suppress the slope in
each forum, leaving almost no room for variation across forums.

The slope feature $\tilde{\beta}$ is then defined as $\tilde{\beta}=\arg\min_{\beta\in(0,\infty)}ERR(\beta)$.
This feature can be estimated with least squares fitting. The slope
feature $\tilde{\beta}$ is essentially a tangent of the angle between
the fitted line and the x-axis. Note that this is not the same as
taking the mean of the slopes of individual path fits.

Our slope feature $\tilde{\beta}$ can be interpreted in terms of
the reciprocity of communication. For example, the soccer forum has
a slope that is closer to one than the slope of the weather forum
(Figure \ref{F:allforums}). This indicates that users in the soccer
forum tend to be involved in balanced conversations, and receive replies
to almost all posts, whereas in the weather forum, users tend to post
more than they receive replies.

Our second feature characterizes the spread or variation in user paths.
This feature is called the spread feature and it is defined as the
fitting error $ERR(\tilde{\beta})$ by Equation \ref{E:err}. For
example, the poker forum has a larger fitting error $ERR(\tilde{\beta})$
than the soccer forum, and this indicates that there is greater diversity
of user paths in the poker forum. This can be visually confirmed in
Figure \ref{F:allforums}.

\subsubsection*{Slope, Baseline and Offset}

The slope feature reflects the averaged reciprocity of communication
in a forum, whereas on the individual level, the slope reflects {}``one-to-many''
reciprocity where a user contributes to a public domain, but can expect
some reward from the public domain.%
\footnote{There is also a hidden reward, because even without receiving replies,
reading the forum can be a rewarding experience. However, the reading
counts are usually difficult to account for numerically. We do not
have these counts for our datasets.%
}

On the level of the forum, the slope feature can give an overall impression
of the type of communication in forums, and we plot the forum slope
in our visualizations. However, the slope can be split down further
in two components. The first component is the overall ratio of the
number of replies to the number of posts in the forum during the period
of observation. We call this ratio a baseline feature $\beta_{0}=\#replies/\#posts$.
Recall that a post can also be a thread starter, in which case, it
is not a reply to any other post. Therefore, our baseline feature
effectively reflects the average number of threads per forum compared
to the average thread length. The baseline feature is an aggregated
statistic over the whole period of observation, and it is not sensitive
to the shapes of the user paths as they evolve over time.

On the other hand, the slope feature itself is sensitive to the ordering
of events in paths. Therefore, our second component of the slope is
the offset feature, which is a signed difference $d=\tilde{\beta}-\beta_{0}$.
The offset feature shows the variance in conversation structure that
cannot be captured by the baseline feature.

Consider two fragments of user paths, one with five posts and five
replies and another one with five replies and five posts. The baseline
for both paths is the same, whereas the slopes and hence the offsets
differ. The offset is negative for the first path and positive for
the second. This reflects the fact that the second user has always
been in a position of a {}``response-attractor'': at each point
of his career he has more or the same number of replies as posts,
whereas the first user has never been in that position.

\subsubsection*{Number of Users and Length of Paths}

Other important properties of a forum are the number of participating
users $N=|U|$ (this is called the size feature) and the average length
of user paths $L_{avg}=\frac{1}{N}\sum_{u\in U}L(u)$ (called the
length feature).

$N$ can reflect the popularity or the generality of a forum topic,
and $L_{avg}$ can reflect the relative persistence of users in the
forum. For example, within the same period of observation, the martial
arts forum has a much larger $L_{avg}$ than the accommodation forum,
which indicates that users in the martial arts forum are more persistent
(tend to post more) whereas users in the accommodation forum tend
to leave after making a small number of posts.\\

We note that in Section \ref{S.5.1:consistent} we studied the patterns
in individual user paths and we omitted short paths. However, in the
calculation of our forum features, we used all paths.

We summarize our forum features for nine \emph{Boards.ie} forums in
Table \ref{T:features}. We compute the four features as we described
above, with the exception of one user from the gigs forum (see Supplementary
Information 3). Note that the baseline accounts for a large part,
but not all of the variation in the forum slopes. The Pearson correlation
between the slope and baseline features is $0.67$.

\begin{table}[h]
\caption{\label{T:features}Summary of forum features for nine forums from
\emph{Boards.ie}}

\centering{}%
\begin{tabular}{ccccccc}
\hline 
\textbf{Name}  & \textbf{Size}  & \textbf{Length}  & \textbf{Slope}  & \textbf{Base} & \textbf{Offset} & \textbf{Spread} \tabularnewline
\hline 
politics  & 881  & 28.89  & 0.98  & 0.97 & +0.01 & 11.55 \tabularnewline
soccer  & 243  & 117.36  & 0.95  & 0.97 & -0.01 & 13.56 \tabularnewline
poker  & 610  & 110.39  & 0.89  & 0.92 & -0.03 & 38.59 \tabularnewline
martial  & 520  & 66.38  & 0.91  & 0.90 & +0.01 & 21.57 \tabularnewline
personal  & 1952  & 21.99  & 0.79  & 0.92 & -0.13 & 9.86 \tabularnewline
accomm.  & 787  & 13.93  & 0.81  & 0.86 & -0.05 & 7.85 \tabularnewline
gigs  & 1151  & 7.32  & 0.84  & 0.80 & +0.04 & 2.03\tabularnewline
weather  & 186  & 9.98  & 0.78  & 0.86 & -0.08 & 1.87 \tabularnewline
develop.  & 355  & 11.86  & 0.86  & 0.87 & -0.01 & 3.93 \tabularnewline
\hline 
\end{tabular}
\end{table}

\subsection{Automated Categorization of Forums\label{S.6.2:categorization}}

We next use our forum features as the basis for performing a hierarchical
clustering of forums. We then show how this clustering allows us (i)
to validate the proposed forum features, and (ii) to demonstrate the
utility of the features.

Prior to the clustering, we normalize each feature by its variance,
because otherwise some feature values are orders of magnitude larger
than others, and hence have a much higher impact on the clustering.
The slope feature was not used in the clustering, because it is included
as the baseline and offset features. The baseline and offset features
were normalized by the variance of the baseline feature, because these
two features are on the same scale with dominating baseline values.
There is no need to align the features to zero mean, because we use
Euclidean distance for the similarity between points, and hence only
the difference between values matters.

We do not have a predefined number of clusters, and therefore we use
hierarchical clustering. We use agglomerative hierarchical clustering
with Ward's linkage method and the Euclidean distance. We also tried
other approaches such as single linkage or average linkage, but observed
similar groupings. We report the clustering where the grouping is
most prominent. We use Euclidean distance because it is widely used
and it is compatible with different linkage strategies. We do not
use cosine similarity, since this measure is not sensitive to the
magnitude of feature vectors. We discuss our clustering results below.

\subsubsection*{Hierarchical Clustering}

\begin{figure}[h]
\begin{centering}
\includegraphics[scale=0.6]{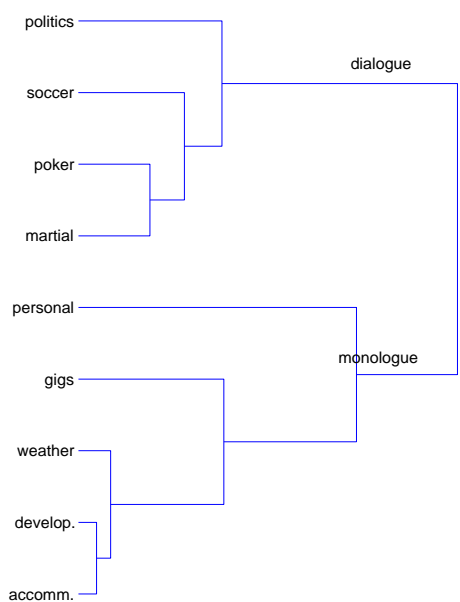} 
\par\end{centering}

\caption{\label{F:dendr}Hierarchical clustering of forums. A prominent feature
of the clustering is the presence of two distinct groups that we labeled
with the terms monologue and dialogue. The users in monologue forums
tend to post more than they receive replies, whereas the users in
dialogue forums tend to have more balanced conversations in terms
of the numbers of posts and replies.}
\end{figure}

The resulting hierarchical clustering is shown in Figure \ref{F:dendr}.
A prominent feature of the clustering is the presence of two distinct
groups that we have labeled with the terms monologue and dialogue.
The forums in the monologue branch tend to have slopes closer to the
x-axis than the forums from the dialogue branch. That is, the users
in monologue forums tend to post more than they receive replies, whereas
the users in dialogue forums tend to have more balanced conversations
in terms of the numbers of posts and replies. This justifies the choice
of labels for the two branches. Alternatively, monologue forums can
be viewed as announcement or classified forums, whereas dialogue forums
can be viewed as discussion forums.

The personal forum stands apart, mostly because it has the largest
offset between the baseline and slope. In the visualization, this
fact is reflected with the presence of user paths that deviate significantly
from the baseline. Furthermore, in each branch, the politics and gigs
forum stand apart because they have the largest number of users in
the corresponding group.

The automated categorization of forums demonstrates the utility of
our forum features. The categorization can be used by forum administrators
to select different management policies for different categories of
forums. Further, the categorization of forums can inform the structural
design of forum websites.

It is hard to define the ground truth clustering for forums, and researchers
often have to appeal to intuitive categorizations of forum names \cite{chan2010decomposing,morzy2010analysis}.
Therefore, we validate our clustering, and hence our forum features,
indirectly. We note that the hierarchical clustering aligns with a
possible manual clustering of visualizations (Figure \ref{F:allforums}),
as well as with a possible intuitive grouping of forums (e.g., the
soccer and martial arts forums group together). Furthermore, the interpretable
clustering of forums that we were able to produce demonstrates potential
value of our forum features.

Moreover, given the explanatory nature of unsupervised clustering,
a misalignment with respect to a hypothetical ground truth or with
respect to the intuition is not necessarily a negative result. We
note that our clustering reveals new information. For example, it
might not be obvious beforehand that the personal forum should be
put in the monologue category, or that the accommodation and development
forums comprise the most similar pair of forums. We further highlight
the novelty of our clustering by comparing it to a result from previous
work.

\subsubsection*{Comparison with Previous Work}

As mentioned in Section \ref{S.1:intro}, we depart from past work
on roles by exploring the stability of user patterns over time, rather
than assuming a stable behavioral role within an ecology of participatory
roles. Moreover, we provide simplified statistics for assessing roles
compared to factor analytic approaches \cite{chan2010decomposing,gonzalez2010structure}
or multi-metric assessments \cite{welser2007visualizing}.

Using this technique we also gained new insights on old data: The
categorization of \emph{Boards.ie} by Chan et al. \cite{chan2010decomposing}
produced an interesting if unexpected series of similarities across
boards. For example, `Politics' and `Accommodation' appeared very
similar in role composition. Given the unexpected nature of these
findings, it made theorization difficult. Also, past work on thread
structure has shown that discussion structure tends to map on to expected
notions of discussion intensity \cite{gonzalez2010structure}. To
this end, the hierarchical clustering we have shown in this paper
has face validity, despite requiring no knowledge of the content of
the forums. This is not to undermine earlier work, but to point towards
techniques that reinforce existing expected similarities and draw
upon behaviors that we consider to be especially salient. We assert
that persistence in the board and tendency to reciprocate are salient
features. The politics forum consists of highly persistent users and
reciprocal communication, whereas and the accommodation forum exhibits
less reciprocal communication and ephemeral users. Additionally, looking
at the role composition of the personal forum, one can suggest that
most people are quiet, whereas our study reveals that there are a
notable minority of users who are able to consistently generate a
lot of attention, probably as a means of garnering social support.

Rather than stating that only one of the methods is correct, we conclude
that our approach is complementary to previous work, as it provides
an alternative view on the forums. However, three of the advantages
of our method are its computational simplicity, its focus on metrics
known to be of relevance to forum participants, and its capacity to
provide both micro-level statistics about users and global statistics
on interactions over time.

\subsubsection*{Alternative Clustering: Baseline Only}

\begin{figure}[h]
\begin{centering}
\includegraphics[scale=0.6]{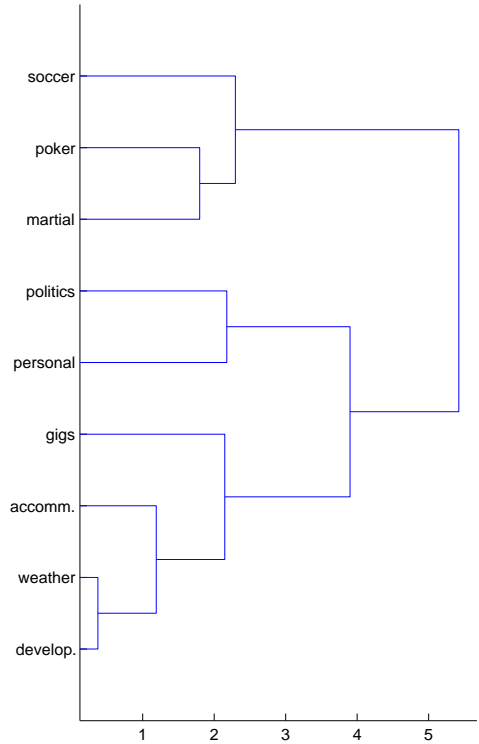} 
\par\end{centering}

\caption{\label{F:alt-dendr}Alternative hierarchical clustering that ignores
the difference between the baseline and slope features. This clustering
differs from the main clustering. For example, the alternative clustering
is less balanced.}
\end{figure}

Recall that the baseline feature accounts for a large part of variation
of the slope feature. In this section, we study the effect of replacing
the slope feature with the baseline feature. We produce a hierarchical
clustering as described above but without the offset feature (Figure
\ref{F:alt-dendr}). We observe that, despite the baseline and slope
features being correlated, the clustering results differ.

A prominent effect from ignoring the difference between the baseline
and slope features, is that the personal and politics forums appear
to be relatively similar and form a cluster. However, as can be seen
in Figure \ref{F:allforums} (see also Section 3 in Supplementary
Information) a certain portion of paths in the personal forum asymmetrically
diverges from the baseline, whereas the paths in the politics forum
tend to follow the baseline more closely. The baseline feature alone
was not able to capture this specific trend, and we suggest using
both the baseline and the offset features as they encompass more information
about the user behavior.

\subsection{Influential Users in the Personal Forum}

We note that the personal forum has the highest offset compared to
other \emph{Boards.ie} forums. Therefore we test whether this is an
effect of a few particular users that have a high influence on the
slope feature or a pattern that is supported by the majority of the
users in the forum. We define the \emph{slope influence} of a user
as a signed difference between the slope estimated with the user path
included and the slope estimated without the path. Interestingly,
we find that there are no exceptionally influential users: $99\%$
of the individual slope influences fall in the interval $[-0.002;0.001]$.

Although visualizations of the personal forum (Figure \ref{F:allforums}
and Section 3 in Supplementary Information) show that some users deviate
from the slope, excluding individual users does not have a strong
effect on the estimation of the slope. We conclude that the difference
between the baseline and the slope is not a result of outliers, but
the pattern that arises from the behavior of the majority of users.
This difference is captured by our features, which provides an additional
justification for our methodology.

\section{Conclusions\label{S.8:concl}}

Forum data provide numerous potential avenues for analysis and visualization,
many of which we have mentioned above. Our contribution to this area
has been to focus on mechanisms that past work have considered salient
(reciprocity and persistence) and that can be employed in a time decoupling
approach. In our approach, we decoupled timing information into user
paths and timing vectors, and studied these parts separately. We constructed
user paths using two new features: the number of a user's posts and
the number of replies to the user over time.

We studied the paths of over $30,000$ users from four forum websites,
and found that users tend to exhibit consistency in communication
over time. We proposed to model the user paths with coin toss and
stickiness models and found that in certain forums, users tended to
be more consistent than one would expect for a random user modeled
with the coin toss model. We also found zones in the feature space
where users were unlikely to be located. Furthermore, we studied the
timing vectors for these users and observed bursty behavior, a well-known
phenomenon in other domains (see for example \cite{barabasi2005origin}
and references therein).

We used our results to represent forums in terms of the temporal behavior
of users, and validated our forum features with a hierarchical clustering
of forums. We proposed a number of applications for our results, such
as predicting whether a user is likely to be involved in conversations
in the future based on past observations, and identifying the normative
behavior for a given forum.

In addition to our main results, we wish to emphasize the flexibility
and generality of our approach, since it can be applied to other domains,
such as email or phone communications. This leads to potential future
work in expanding this technique. One expansion would be to consider
different media, such as a call graph, where posts and replies are
replaced with patterns of initiation between calling partners. In
this case, it is known that initiation patterns vary by media \cite{carrasco2008agency},
but exact values and a role analysis is yet to be done. Another extension
would consider alternate feature spaces, such as posts that are weighted
(such as Amazon's `useful' score or Reddit and more recently YouTube's
upvotes/downvotes scores). The latter case might provide many interesting
revelations about trolling and flaming given that a user's score may
fall below zero. By contrast, in the current case, all replies increase
the value of the user's score.

Finally, we believe our technique provides a powerful way to look
not just at the differences between users, but at the collective evolution
of communities. Trending slopes over time may provide an indicator
for moderators about the success of their community, or be employed
as real-time statistics for site owners about the health of their
site and the intensity of discussion therein.

\subsection*{Acknowledgments}

We would like to thank Václav Belák from Digital Enterprise Research
Institute for preparing the \emph{TiddlyWiki} dataset, and Lee Jensen
from Ancestry.com Inc. for help in obtaining the \emph{Ancestry.com}
dataset.

This work is partially supported by the Science Foundation Ireland
(SFI) under CLIQUE Strategic Cluster, grant number 08/SRC/I1407. This
work is partially supported by National ICT Australia (NICTA). NICTA
is founded by the Australian Government's Backing Australia's Ability
initiative, in part through the Australian Research Council.

\bibliographystyle{spmpsci}
\bibliography{victoria,library}

\end{document}